# Modes of Collaboration in Modern Science – Beyond Power Laws and Preferential Attachment

Staša Milojević

School of Library and Information Science, Indiana University, Bloomington 47405-1901. E-mail: smilojev@indiana.edu

The goal of the study is to determine the underlying processes leading to the observed collaborator distribution in modern scientific fields, with special attention to non-power law behavior. Nanoscience is used as a case study of a modern interdisciplinary field, and its coauthorship network for 2000-04 period is constructed from NanoBank database. We find three collaboration modes that correspond to three distinct ranges in the distribution of collaborators: (1) for authors with fewer than 20 collaborators (the majority) preferential attachment does not hold and they form a log-normal "hook" instead of a power law, (2) authors with more than 20 collaborators benefit from preferential attachment and form a power law tail, and (3) authors with between 250 and 800 collaborators are more frequent than expected because of the hyperauthorship practices in certain subfields.

## Introduction

Neither collaboration nor coauthorship, its most studied indicator, is well understood phenomenon despite an intense interest in the topic. Furthermore, the focus of most studies is typically on highly connected scientists, i.e. the stars, and the processes leading to star formation. However, most scientists are not stars and yet there has been little effort to understand coauthorship practices and collaborative behavior of this majority. This study is a step in that direction. By exploring a very active emerging interdisciplinary field as a case study, this paper is an attempt to understand modes of collaboration other than those of highly successful individuals.

Collaboration, both formal and informal is one of the defining features of modern science, and yet the concept is difficult to define. After reviewing a number of definitions, Hara et al. (2003) determined that they share two common elements: "working together for a common goal and sharing of knowledge" (p. 853). However, a number of other definitions emphasize the importance of collaboration as a social process (Bordons & Gomez, 2000; Shrum, Genuth, & Chompalov, 2007) or stress that it takes place in a specific social context (Sonnenwald, 2007). Collaboration has been extensively studied since the 1960s (Beaver, 2001; Beaver & Rosen, 1978, 1979a, 1979b). The most commonly used methods are bibliometrics (Bordons & Gomez, 2000; Glänzel, 2002; Glänzel & Schubert, 2004), social network analysis/network science (Barabási et al., 2002; Kretschmer, 1997; Newman, 2001b, 2004; Wagner, 2008; Wagner & Leydesdorff, 2005), qualitative methods of observation and interviews (Hara et al., 2003; Shrum et al., 2007) and surveys (Birnholtz, 2006; Lee & Bozeman, 2005). Coauthorship, a form of association in which two or more scientists jointly report their research results on some topic, is the most visible indicator of collaboration and has thus been frequently used to measure collaborative activity, especially in bibliometric and network analysis studies. Bibliometric studies of coauthorship have focused mainly on the effects of collaboration on scientific activity as well as on organizational and institutional aspects of collaboration, looking at authors, research centers/institutions and countries as units of analysis. Network studies, on the other hand, have focused primarily on reconstructing the actual collaboration networks and understanding the underlying structures, mechanisms and processes leading to the observed structures. Coauthorship is not the same as collaboration, and therefore it can be used only as a partial indicator of collaboration (Katz & Martin, 1997; Smith & Katz, 2000). That is to say not all collaborations lead to coauthored papers, and not all coauthorships are the result of research collaboration in a strict sense (Luukkonen, Persson, & Sivertsen, 1992; Melin & Persson, 1996; Subramanyam, 1983). While aware of these caveats, in this work we will follow other studies of collaboration through coauthorship networks and consider coauthorship to imply collaboration. The only (yet important) distinction is purely technical: in the rest of this paper the term *collaborators* is used to represent the totality of authors (or, more precisely, distinct author *names*, as discussed below) with whom a given author has coauthored papers over some time period, while the term *coauthors* is used to mean authors that appear together on a single paper.

It is only recently, with the advancement of computer power and digital data availability, that large-scale network studies of collaboration through coauthorship have become possible. A number of these studies used collaboration networks[1] to study network dynamics (Barabási & Albert, 1999; Barabási et al., 2002; Farkas et al., 2002; Newman, 2001a). These studies reveal the existence of small-world and scale-free network topologies and preferential attachment as a structuring factor. Small-world networks are networks with high local clustering and short global path lengths. Preferential attachment in the context of networks means that the well-connected nodes are more likely to attract new links. The distribution of the number of collaborators is an important property for determining the topology of a network and processes leading to its evolution. Since the number of collaborators each author has in a network is simply the degree of a node, the distribution of the number of collaborators is the same as the node degree distribution. Networks in which the degree distribution is typically right-skewed, meaning that the majority of nodes have less-than-average degree and that a small fraction of hubs are many times better connected than the average, are scale-free networks and follow a power law functional form (Börner, Sanyal, & Vespignani, 2007; Newman, 2003; Watts, 2004a).

Moody (2004) tied the model of scale-free networks to the existence of scientific stars, or high-status scientists. The observation that there is a "star system" in science, i.e. that a small number of researchers, or scientific stars, shape the discipline by using their central position in a field to diffuse their ideas, dates back to the research on stratification in science (Allison, Krauze, & Long, 1982; Cole & Cole, 1973; Crane, 1972; Merton, 1968; Zuckerman, 1977). Star scientists are centrally positioned because they attract disproportionate levels of funding, students and collaborators.

One way to explain the large inequality in the number of collaborators is through a process of preferential attachment (Barabási & Albert, 1999). The process by which some resource is distributed among a number of individuals in such a way that those who have a lot receive more than those who do not, is also known as: the "Matthew effect" (Merton, 1968, 1988), cumulative advantage (Allison et al., 1982; Price, 1976) or accumulative advantage (Cole & Cole, 1973). Following Newman (2001a) and Barabási et al. (2002) one can discriminate between three types of preferential attachment in collaboration networks:

(a) Papers appear with author names that have not published before (*new* authors). If these papers contain more than one author and some have published before, new authors will join the collaboration network by forming connections with *old* (i.e., *existing*) authors. Preferential attachment states that new authors attach with higher probability to those old authors who already have many connections (collaborators). If there is a linear relation between the probability of a new connection and the node degree (the number of previous collaborators) this process alone leads to a power law degree distribution (Barabási & Albert, 1999).

(b) As a network grows one may identify not only connections with new actors, but also new connections between authors who were already present in the collaboration network but did not collaborate with each other before (they did collaborate with certain other authors, which is why they are in the network). According to Newman (2001a), for existing authors who have not collaborated before, preferential attachment means that the probability of collaboration depends on the number of *common* (i.e., mutual) collaborators. The higher the number of common collaborators of two authors, the probability that they will form a collaboration themselves would be higher. This form of preferential attachment is also supposed to explain high clustering coefficients. Barabási et al. (2002) interpret preferential attachment between existing authors in a different way. For them, the probability of a collaboration between existing authors is proportional to the number of collaborators that each of the authors has, but these do not need to be just mutual collaborators. Dorogovtsev and Mendes (2000) consider this an extension of the concept of preferential attachment of new authors.

---

[1] These networks are usually called coauthorship networks, but here they will often be called *collaboration networks* to be consistent with the distinction between our definitions of collaborators and coauthors.

(c) Finally, for authors who have already collaborated, preferential attachment would mean that the probability of another collaboration depends on the number of times they have collaborated in the past. Note that this type of activity does not change the topology of a network because all nodes and connections remain the same.

Network studies of collaboration through coauthorship using large-scale datasets were used to examine a number of scientific fields and disciplines: Astrophysics, Polymer science, Geophysics, Soil Science, Virology, Medicine, Physics, Computer science, Mathematics and Neuroscience. However, most of these studies focused only on the small-world and scale-free properties of the resulting networks. They did not, with the exception of Wagner and Leydesdorff (2005), explore processes other than preferential attachment that would lead to the observed distributions of collaborators. In other words, most previous studies explored only the power law regime[2] of the collaborator distribution (the part that appears as a straight line in log-log plots) and dismissed any other features of collaborator distribution even when they were highly statistically robust, and not merely statistical fluctuations. Wagner and Leydesdorff, in a smaller scale study (covering one year and 19,147 articles), notice that for authors with a small number of collaborators the number of such authors falls below what is expected from a power law fit to the tail and often has a peak. They call this deviation from the power law distribution a "hook". They suggest that the regime involving authors with small number of collaborators is the result of the arrival of newcomers to the field, while the regime involving authors with many collaborators is the result of continuing authors and those that leave the network.

The goal of this study is to investigate the distribution of collaborators in coauthorship networks in an emerging, highly interdisciplinary field, paying special attention to non-power law behavior and its causes. Given that the rate of coauthorship changes over time and across disciplines and given, too, the effort that science policy makers are putting into supporting collaborative teams, looking at a new field can give us a better sense of coauthoring practices in those highly active and successful fields today. An obvious candidate for such a study is nanoscience/nanotechnology[3].

There is a general view that nanoscience presents a modern scientific field. Broadly defined, nanotechnology is a research field studying objects that have a size or structure of 1-100 nanometer (a nanometer is one millionth of a millimeter). It is an exceptionally good choice as an example of a modern scientific field for a number of reasons. It has variously been described as a "new frontier" (Barben, Fisher, Selin, & Guston, 2008), an "emergent field" (Wajcman, 2008), an "emergent, highly interdisciplinary field" (Zucker, Darby, Furner, Liu, & Ma, 2007), a "transdisciplinary research front" (Hayles, 2004) and a "rigorous scientific field" with "many signs of protodisciplinarity" (Milburn, 2004). Nanotechnology blurs the boundaries between research and application and between science and engineering, having formed at the intersection of the unprecedented number of fields in science and engineering. It has mobilized governments, industry, entrepreneurs, venture capitalists, and nongovernmental organizations (Berube, 2006; Grodal, 2007).

In this study we first analyze the distribution of collaborators in a five-year period (2000-04) within nanoscience and processes leading to such distribution using social network analysis. We then examine three distinct behaviors: authors with a small number of collaborators, authors with a large number of collaborators, and authors with an extremely large number of collaborators. We then test whether processes of preferential attachment (all three types) are present and to what extent. We finally look at six different nanoscience subfields and examine the distribution of collaborators within each.

---

[2] The word regime is used in this paper to connote the characteristic behavior of the collaboration process.

[3] Terms nanoscience and nanotechnology are used interchangeably throughout this study.

**Methods**

The principal source of data for this study is the article section of NanoBank database (version Beta 1, released on May 2007) (Zucker & Darby, 2007). NanoBank is a digital library of bibliographic data on articles, patents and grants in the field of nanotechnology (Zucker & Darby, 2005). A set of nanotechnology-related articles in NanoBank has been selected from the *Science Citation Index Expanded, Social Sciences Citation Index,* and *Arts and Humanities Citation Index* produced by Thomson Scientific (now Thomson Reuters). Two separate methods of selecting nano-related documents have been used in the creation of NanoBank (Zucker et al., 2007): (1) selecting articles that contain some of the 379 terms identified by subject specialists as being "nano-specific", and (2) selecting articles based on a probabilistic procedure for the automatic identification of terms. The database covers a 35-year period (1970-2004). However, the focus of this paper is mainly on the latest 5-year period (2000-04), i.e., present-day nanoscience. In 2000-04, according to NanoBank, 294,456 authors published 270,135 articles in 4,792 journals (see Table 1). Given its coverage, large size and solid construction methods, NanoBank provides what appears to be a rather comprehensive representation of nanotechnology, and thus a rich source for examining collaboration patterns.

The problem of distinguishing individual authors based on names as given in large datasets has already been acknowledged as one difficulty in obtaining accurate results in studies involving authors (Moody, 2004; Newman, 2004). This problem is present in this study as well. Namely, it is impossible to determine with absolute accuracy how many different authors there are in the dataset. One can instead say how many distinct *names* appear in the database. A single author may report his/her name differently on different papers, which leads to the problem of collocating synonyms. On the other hand, two authors may have the same name, which leads to the problem of disambiguating homonyms. Two most common ways of identifying authors in large datasets are: using the last name plus any initials to represent separate authors and using the last name plus only the first initial to separate authors. These two schemes have been called by Newman (2004) *all initials* and *first initial*. This study follows the more accepted practice of taking into account all initials, but in one case where this choice may be pertinent we also explore the effect of homonyms.

Using all articles from NanoBank published between 2000 and 2004 we constructed a non-directed collaboration network where the nodes are authors who have collaborated and edges are coauthoring events. Most of the authors are part of the collaboration network, since in 2000-04 only 1% of authors have published exclusively single-author papers. This shows that collaboration in nanotechnology research has become the norm and networks can be used to analyze the behavior of the majority of authors.

Data on nanotechnology subfields are not part of the NanoBank database. To supplement this information the 2005 *ISI Journal Citation Reports (JCR)* were used to obtain subject categories of journals, which were used as an approximation to determine authors' research subfields. Using *ISI JCR* 42 nanotechnology subfields were identified (Milojević, 2009). Subfields were assigned to authors based on the subfield most prevalent in each author's oeuvre. In cases where there were two or more most frequent subfields with tied counts, one of these subfields was assigned to author randomly (this was the case for only 18% of all authors).

| Number of | Total | 1995-1999 | 2000-2004 |
|---|---|---|---|
| Journals | 8,300 | 4,255 | 4,792 |
| Articles | 580,710 | 172,583 | 270,135 |
| Authors | 466,603 | 208,191 | 294,456 |

Table 1. Number of journals, articles and authors in NanoBank.

## Results

*Distribution of the number of collaborators per author (node degree)*

The distribution of the number of collaborators is an important indicator of the structure of a collaboration network and of the processes that produce that structure. Since the number of collaborators that each author has in a network is simply the degree of a node, the distribution of the number of collaborators is the same as the node degree distribution.

On average, each author collaborates with 15.8 authors. The distribution[4] of the number of collaborators of a given author exhibits three distinct features: the "hook" (where distribution curves for small values of $k$, where $k$ is the number of collaborators); the power law tail; and the anomalous peak in the tail (see Figure 1). Presence of the hook leads to one important feature of the distribution – that it is *not* entirely scale-free. Most authors have three collaborators, and the existence of this peak gives scale to the distribution. The number of authors with an increasing number of collaborators falls monotonically in a power law fashion until a second peak occurs for authors with between 250 and 800 collaborators.

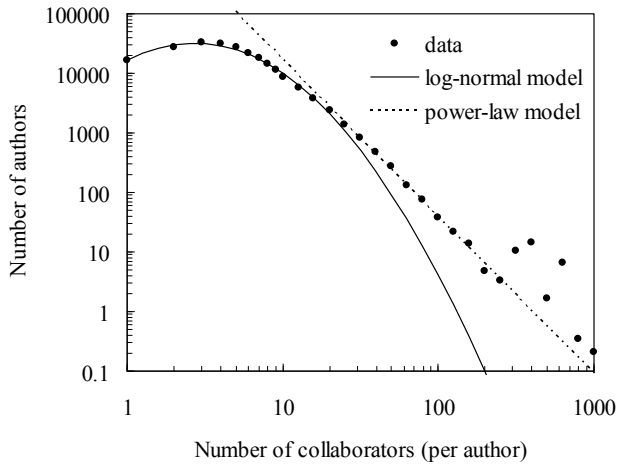

**Figure 1.** The log-normal and linear models fitted to the distribution of the number of collaborators of a given author (based on NanoBank database and restricted to the period 2000-04). Because of the logarithmic binning the number of authors (actually a probability) can have values less than one.

The part of the distribution below 20 collaborators follows very closely a log-normal distribution[5]. A large majority of authors belongs in this part of the distribution (88% of collaborating authors). Log-normal distribution is similar to the normal (or Gaussian) distribution except that the distribution over $k$ is logarithmic and not linear. Like a normal distribution, a log-normal distribution can have a maximum value. Figure 1 shows the log-normal fit to degree distribution for authors with up to 20 collaborators. Apparently, the log-normal model is a good description for this part of the distribution and is given by:

$$\log y = -1.5755 (\log k)^2 + 1.3556 \log k + 4.2186,$$
where $\log k < 1.4$

where $k$ is the number of collaborators and $y$ the number of authors with $k$ collaborators. The fitted peak (mode) is at 2.7 collaborators per author. The log-normal distribution stops being a good model for 20 or more collaborators per author. From that point and until we reach an anomalous second peak, the power law, given by the linear fit, is an excellent description of the distribution:

$$\log y = -2.6349 \log k + 6.8594, \text{ where } 1.2 < \log k < 2.5$$

The exponent of 2.63 agrees well with the exponent range from 2 to 3 usually found in collaborative networks (Dorogovtsev & Mendes, 2002). Finally, there is an anomalous peak on top of the power law tail to which no functional form is fitted.

Given that the distribution of the number of collaborators per author exhibits three distinct regimes,

---

[4] At high values of $k$ discrete power law distributions usually start to be dominated by noise, especially since the number of collaborators cannot be smaller than one. To eliminate the effects of noise, for $k \geq 10$ we perform logarithmic binning of the data in bins of 0.1 decades. This procedure allows an unbiased characterization of the distribution much beyond the point where it is possible with unbinned data.

[5] In the same way in which the power law function becomes represented with a straight line in the log-log plot, the log-normal distribution is represented with a *quadratic* function (second order polynomial) in the log-log plot, a curve which looks like a *parabola*.

the nature of each will be examined separately: the anomalous peak, the "hook" and the power law.

*Nature of the anomalous peak*

Two possibilities that might explain what is producing the deviation from the power law model leading to a high number of authors with an extreme number of collaborators are explored. The first is that this is an artifact of one's inability to distinguish different authors due to frequently occurring last names. Since in the analysis different authors who would have the same last name and initials are considered as one, this will boost the number of seeming collaborators, perhaps leading to the anomalous peak.

The entire data base contains 203,197 different last names. However, some appear much more frequently than the others. The 100 most frequent last names account for 5% of all author names, a disproportionally high contribution. Removing authors having these names should eliminate the bulk of the effects associated with name homonyms. However, Figure 2 shows that the removal of the 100 most frequent last names does not lead to the disappearance of the anomalous peak. Instead, this ambiguity apparently affects (to a certain degree) the *slope* of the power law.

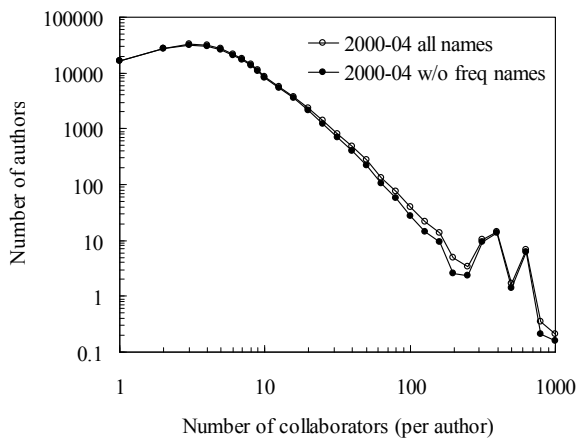

**Figure 2. The distribution of the number of collaborators with 100 most frequent names removed.**

The second possibility for an anomalous number of authors with very large number of collaborators that we explore is that it is the consequence of having a relatively small number of papers with an extremely large number of coauthors. All such coauthors are considered collaborators in the standard network analysis followed here. In nanoscience the very small number of articles written by coauthor teams larger than 20 (see Table 2) shows that it generally does not have coauthorship properties of fields such as high energy physics where "hyperauthorship" is the norm. Cronin (2001, p. 558) introduced the term to refer to "massive coauthorship levels".

| Number of authors per article (x) | Number of papers with x authors | Percent total articles |
|---|---|---|
| 1 | 24823 | 9.2% |
| 2 | 50468 | 18.7% |
| 3 | 54996 | 20.4% |
| 4 | 47645 | 17.7% |
| 5 | 34404 | 12.8% |
| 6 | 23424 | 8.7% |
| 7 | 14163 | 5.3% |
| 8 | 8309 | 3.1% |
| 9 | 4633 | 1.7% |
| 10 | 2681 | 1.0% |
| 11-20 | 3683 | 1.4% |
| 21-50 | 127 | 0.0% |
| >50 | 33 | 0.0% |

**Table 2. Number of authors per article.**

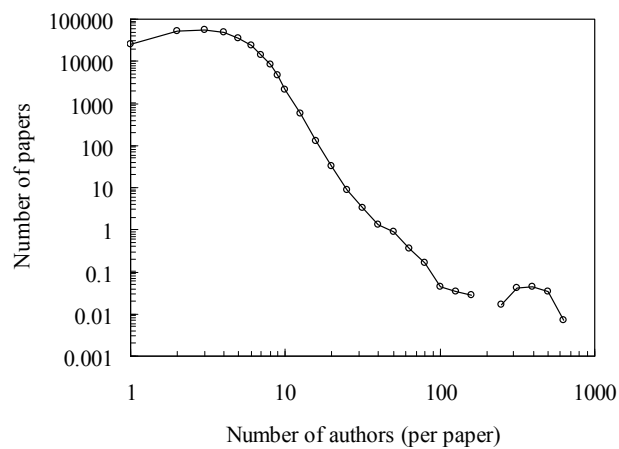

**Figure 3. The distribution of the number of authors per paper (i.e., coauthor team sizes), binned logarithmically.**

At the same time, the results show the presence of a small number of papers that do exhibit properties of "hyperauthorship." Figure 3 shows that the number of papers with more than 200 authors does not follow a distribution set by the papers with smaller coauthor teams, but is significantly higher. In order to check whether hyperauthorship practices have an effect on the existence of the anomalous peak, nanotechnology subfields with such papers were identified. Out of 42 subfields only three have articles with more than 200 coauthors: Multidisciplinary Physics (P1), Physics of Particles and Fluids (P6) and Nuclear Physics (P7). The collaboration network omitting all the articles that belong to these three subfields was then constructed. This leaves 245,096 articles or 83% of the original number, and the collaboration network now contains 274,279 authors or 95% of the original number. The degree distribution of this network is shown in Figure 4 alongside the original distribution. An anomalous peak is no longer present, which is the only respect in which the two distributions differ significantly. The new distribution continues to be a power law in the region where the anomalous peak used to be.

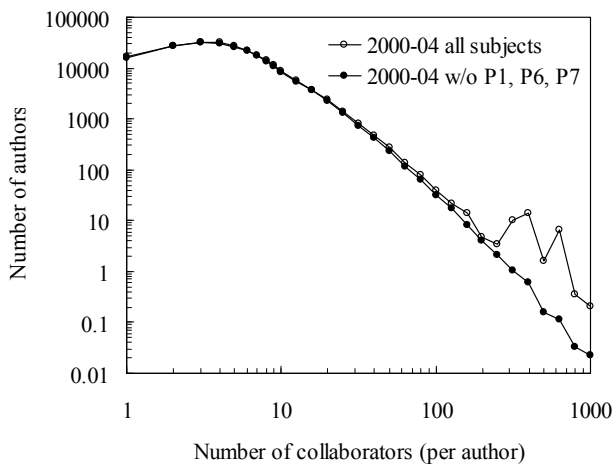

**Figure 4. The distribution of the number of collaborators of a given author with three subfields (P1, P6, P7) removed.**

*Nature of the log-normal "hook"*

Wagner and Leydesdorff (2005) propose that the "hook", i.e., the curving of the degree distribution at low numbers, is the result of the arrival of newcomers and transients to a field. In his study of Social Science, Moody (2004) finds that "those with greater time in the discipline (exposure) are slightly more likely to coauthor" (p. 224). Both studies indicate that time spent in the field may be correlated with collaboration. Therefore, it is here explored how collaborator distribution may depend on factors such as the time an author has spent in the field.

For each of the 289,764 authors in the collaboration network the time spent in the field, defined as the time span between the first nanotechnology article published (as recorded in NanoBank) and the most recent one, was determined. This time period can range from one year (a single year within the 2000-04 period) to 35 years (the full range of NanoBank, 1970-2004). Next, the authors with the same number of years spent in the field were grouped, and for each group the average number of collaborators they had in the current (2000-04) network was computed. This is shown in Figure 5.

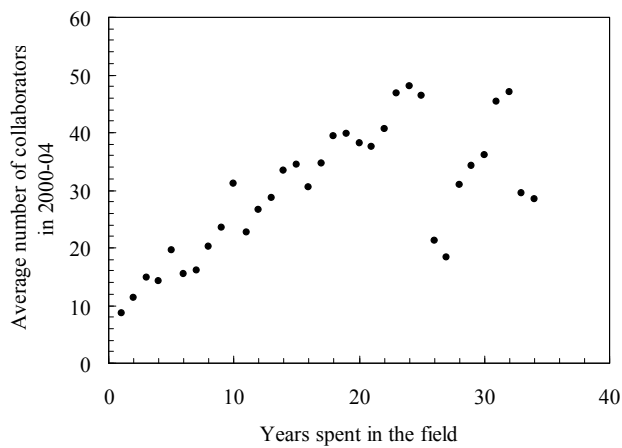

**Figure 5. The relationship between years spent in the field and the average number of collaborators for 2000-2004.**

For authors who have spent up to 25 years in the field there is a clear linear correlation, with each additional year "bringing" 1.6 collaborators on average. While it would not be surprising to find that people who have spent more time in the field have collaborated with more people over their entire careers, here we show that such correlation exists even with the number of collaborators they had during a fixed five-year period. The correlation breaks down for scientists with careers longer than 25 years. It is possible that they have ceased active careers, and are now publishing with only a subset of former collaborators. From this correlation one can see that scientists populating the "hook" of the distribution of number of collaborators (scientists with fewer than 20 collaborators) are more likely to be the ones who have spent 8 or less years in the field (i.e. newcomers and transients). Also, one can see that the average number of collaborators for authors who have just joined the field is already quite high (around 9).

In order to explore the effect of the time spent in the field on the collaborator distribution, two analyses were performed. First, the effect of the gradual removal of authors who have spent less than a certain number of years in the field was examined (Figure 6). The curve with black circles shows the full distribution, i.e., a repeat of the distribution from Figure 1. The curve below it (gray circles) shows the distribution with authors having only one year of experience removed (i.e., it includes all the authors with 2 or more years spent in the field). The one below (open squares) removes all with less than 4 years in the field, followed by less than 8 and less than 14 years (crosses and asterisks). That each curve is below the previous one simply reflects that each is a subset of the preceding. One can see that the effects of removing authors who have been present in the field for the shortest period of time is to lower the part of the distribution where authors with fewer collaborators are to be found (i.e. the "hook"). However, the log-normal character of the distribution is preserved. Moreover, the place where the power law part of the distribution gets established moves to ever higher $k$ values. Also, the authors who have spent more than 14 years in the field appear unlikely to be involved in hyperauthored papers. But most importantly, authors who have spent an increasing number of years in the field tend to have a collaborator distribution that is increasingly dominated by the log-normal regime.

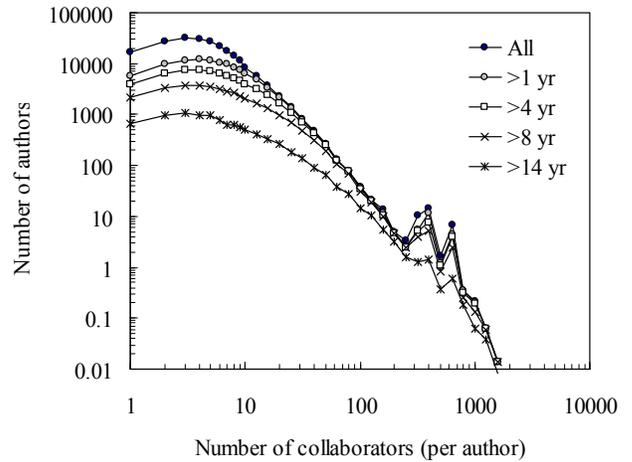

**Figure 6.** The effect on the collaborator distribution of the gradual removal of authors who have spent less than a certain time in the field, i.e., leaving those with more than 1, 4, 8 and 14 years spent in the field.

Next the effect of removing authors having more than a given number of years in the field was examined. In Figure 7 the original distribution is presented as a curve with black dots, the distribution of authors with less than 15 years of experience as a curve with asterisks, followed by the distributions of authors with less than 9, 5 and 2 years (crosses, open squares, gray circles). Now it is the log-normal part of the distribution that stays relatively fixed, but the removal of the authors who have spent a longer time in the field affects the power law part, making it steeper when only the authors with very short presence are kept. The anomalous peak is preserved, signifying that, as the log-normal part, it is mostly built by newcomers to the field and/or the transients, distinction of the two not being possible without future data.

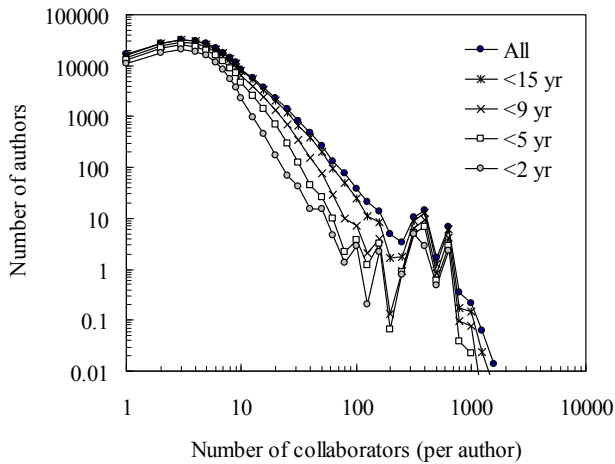

**Figure 7.** The effect on the collaborator distribution of the gradual removal of authors who have spent above certain time in the field (for period 2000-2004).

If the "hook" is a consequence of a separate type of collaboration process, then one would expect the trend of the declining number of authors with very few collaborators to extend to those authors who do not collaborate at all and are absent from the network. Normally, one cannot plot an $k$-value of zero on a log-log plot since log 0 is not defined. Moody (2004) proposes a way to overcome this limitation by adding the number one to values of $k$. The resulting distribution is shown in Figure 8. Apparently, the isolated authors lie at the continuation of the "hook". While valid, this procedure is intended to be used as an illustration, since it cannot be used to obtain a functional form consistent with the log-normal equation given previously.

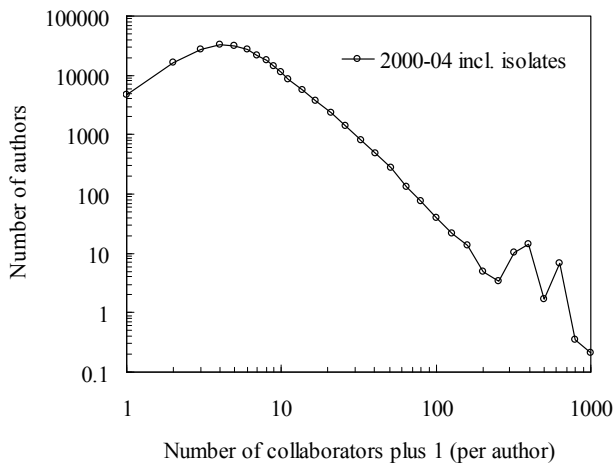

**Figure 8.** The distribution of the number of collaborators of a given author for that includes non-collaborating authors.

*Preferential attachment*

Preferential attachment has been suggested as a process which leads to scale-free network topology in collaboration networks (Barabási & Albert, 1999). Given these results and the fact that a part of collaborator distribution in nanoscience (Figure 1) follows the power law it is next examined whether preferential attachment can be used to explain patterns of collaboration in nanoscience.

To test preferential attachment one needs to distinguish between existing and newly formed collaborations. Collaborations that were present during the five-year period 1998-2002 were considered to be existing. The following two years (2003-2004) were used to give us information on *new* collaborations (whether with new or among the existing collaborators). The old collaboration network, for papers published in 1998-2002, contains 1,850,146 collaborations between 253,285 authors, while the new network, for papers published in 2003-04, has 1,193,683 ties between 184,446 authors. In the next three sections the results of the tests for preferential attachment of new authors to old, among existing authors who have not collaborated, and existing authors who have collaborated will be presented.

*Preferential attachment of New Authors to Old Authors*

To test for preferential attachment of new authors, collaborations that involve a new author and existing author were selected from the new network, noting each time the number of collaborators the existing author had in the old network. If preferential attachment is present, there should be a positive correlation between the average number of new collaborators for the given number of previous collaborators. More specifically, the correlation needs to be *linear* in order for preferential attachment to produce a power law distribution (Barabási & Albert, 1999). The relative probability of collaborating with new authors as a function of the number of existing collaborators is shown in log-log plot in Figure 9. A linear function is represented by a line having a slope of 1. For part of the curve between 10 and 300 collaborators a linear fit with a slope 1.06 is obtained, thus formally confirming linear preferential attachment in this range. However, the relation departs from the linear trend below ~10 collaborators (it becomes shallower), and becomes constant for those with one collaborator, who have a similar chance of acquiring a new collaborator as those with two existing collaborators. The region where linear preferential attachment is not valid again has a counterpart in the degree distribution function (Figure 1). It corresponds to the region where degree distribution follows a log-normal distribution. Finally, for those with more than 300 collaborators preferential attachment no longer holds, corresponding to the region of the anomalous peak in degree distribution.

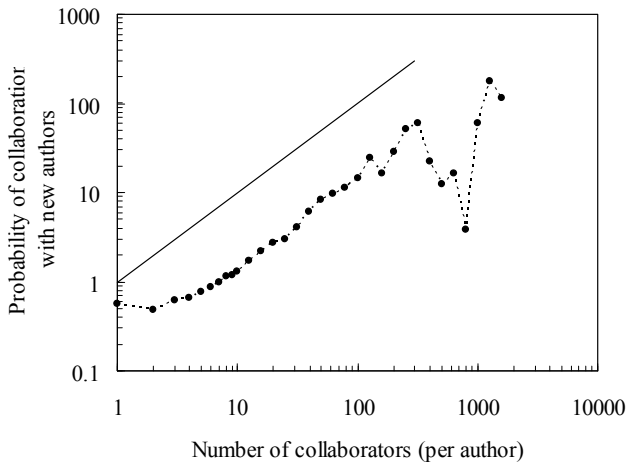

**Figure 9. Relative probability of collaboration with new authors (average number of new collaborators) as a function of the number of existing collaborators in log-log plot. The solid line has a slope of unity and represents linear preferential attachment.**

The analysis is repeated with papers having more than 200 coauthors[6] omitted. Figure 10 shows that the preferential attachment now extends to authors with more than 300 collaborators, all the way to the last data point of 1,000. Apparently, the normal mode of authorship (the authorship that does not exhibit the properties of hyperauthorship) obeys the preferential attachment even for huge number of collaborators (the slope is 1.07 for $k > 10$). This is again consistent with the node degree distribution displaying the power law behavior even for high number of collaborators when fields that produce hyper-authored papers are excluded (Figure 4).

---

[6] In 1998-2002 there are 16 out of 229,195 papers with 200 or more coauthors each.

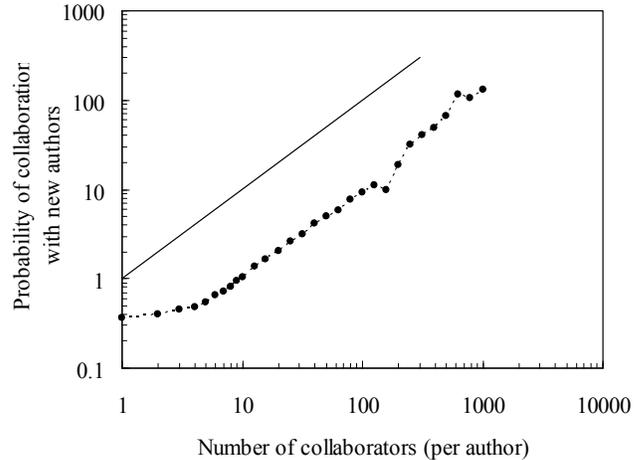

**Figure 10. Relative probability of collaboration with new authors (average number of new collaborators) as a function of the number of existing collaborators in log-log plot with papers with more than 200 authors excluded. The solid line has a slope of unity and represents linear preferential attachment.**

*Preferential attachment among the existing authors who have not collaborated previously*

We next test for the presence of preferential attachment for authors who are not newcomers to the field, but who have not collaborated with each other before. According to Newman, preferential attachment in this case means that the probability of the collaboration will depend on the number of *mutual* collaborators. We select from the "new" network newly formed collaborations and for each such pair find the number of mutual collaborators they had in the "old" network. This number does not automatically give the probability of new collaboration. To determine the probability, one needs to *normalize* this number over what would be the case if collaborations were taking place *randomly*. The probability of new collaboration among existing authors as a function of the number of mutual collaborators is shown in Figure 11.

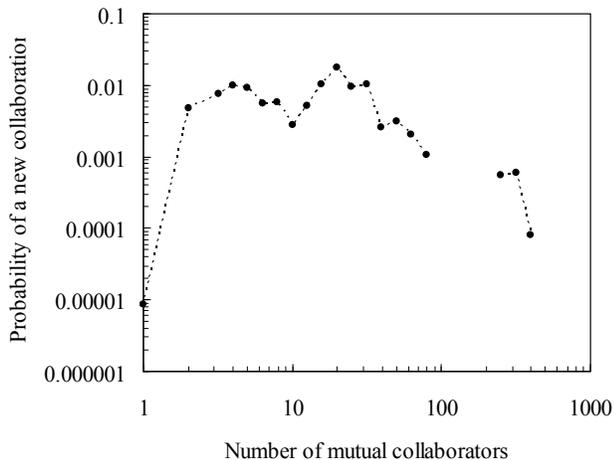
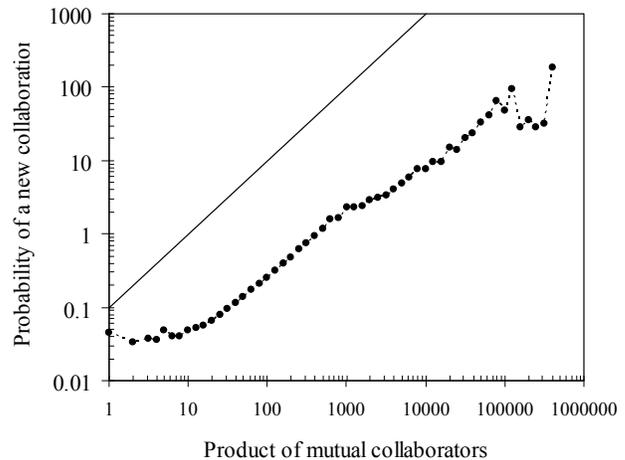

**Figure 11. Probability of new collaboration among existing authors as a function of the number of mutual collaborators. Preferential attachment is not present.**

**Figure 12. Probability of a new collaboration among the existing authors as a function of the product of the number of collaborators of both authors. Linear preferential attachment is shown as a solid line.**

There is no positive correlation of the kind expected in the case of preferential attachment, i.e., the probability does not depend on the number of mutual collaborators. Results are similar when papers with more than 200 authors are excluded (not shown). The lack of evidence for this type of preferential attachment does not mean that the preferential attachment does not exist in some other form for collaboration among the existing authors.

Indeed, another approach to studying collaboration among the existing authors in terms of preferential attachment is that formulated by Barabási et al. (2002). Here, the claim is that the probability of a new collaboration among the existing authors is proportional to the number of collaborators that each author had, regardless of whether they are mutual or not. This is actually an extension of the concept of preferential attachment of new collaborators. We again focus on the most recent networks (old: 1998-2002, new: 2003-2004) and count new collaborations formed in the new network of authors present in the old network, eliminating those who have already collaborated. For each such collaboration we note the product of the total number of collaborators of the first $k_1$ and the second $k_2$ author ($k_1 k_2$). To get the probability, the counts are normalized by the total number of pairs in the network that have $k_1 k_2$ collaborators. Results are presented in Figure 12, a log-log plot with logarithmic binning.

The results are qualitatively similar to those obtained for new authors joining the network (Figure 9). Preferential attachment begins to operate when $k_1 k_2 > 15$ and extends to $k_1 k_2 \sim 120,000$, which corresponds to $\sqrt{k_1 k_2} \sim 300$, i.e., the range of collaborators responsible for the anomalous peak. Below $k_1 k_2 = 15$ preferential attachment is not present, similarly as in the case of new authors. Overall, the slope is slightly shallower than 1 (shown as the solid line) with a value of 0.77 between $20 < k_1 k_2 < 120,000$. The trend appears to have a break at $k_1 k_2 \sim 1000$, the reasons for which are explained in the next paragraph. The slope is closer to one, with a value of 0.91, when the fit is restricted to $20 < k_1 k_2 < 1000$.

Preferential attachment when hyper-authored papers (those with more than 200 authors) are excluded is shown in Figure 13. The results differ from the original ones most prominently when the product of the mutual number of collaborators exceeds 1,000. New figure does not have a break there but continues with the same trend. This trend is now much closer to linear preferential attachment with the slope of 1.06 for the entire range from 20 to 100,000.

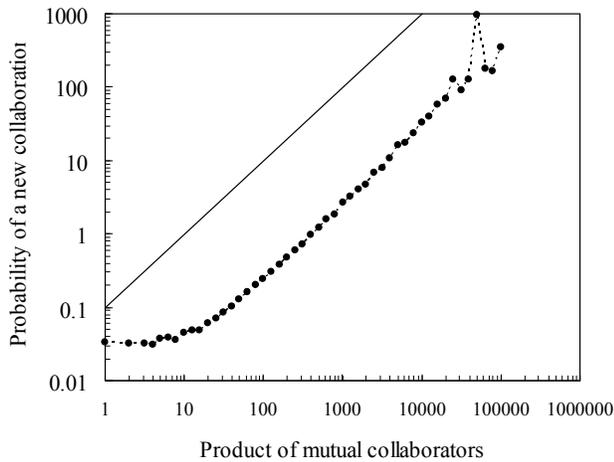

**Figure 13.** Probability of a new collaboration among the existing authors as a function of the product of the number of collaborators of both authors with papers with more than 200 authors excluded. Linear preferential attachment is shown as a solid line.

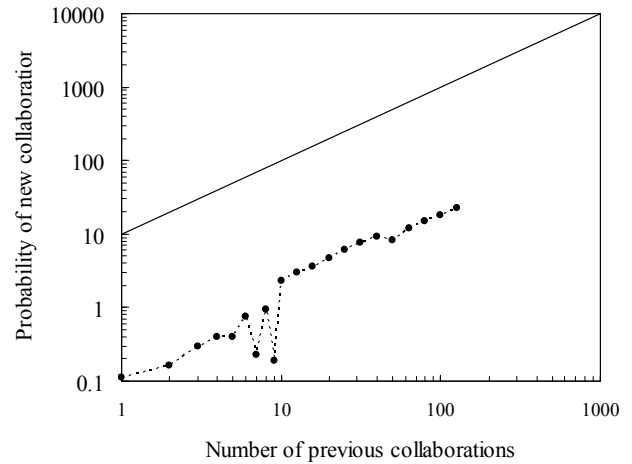

**Figure 14.** Probability of a new collaboration among the authors who have collaborated as a function of the number of previous collaborations. Linear preferential attachment is shown as a solid line.

*Preferential attachment among the existing authors who have collaborated*

According to Newman (2001a), the third type of preferential attachment implies that the probability of new instances of collaboration among the authors who previously collaborated increases as does the number of previous instances of collaboration. This type of activity does not modify the network topology because the two authors are already connected. However, if we take the number of instances of collaboration as edge weight, than this sort of activity would increase the weight.

We again focus on the most recent networks (old: 1998-2002; new: 2003-2004) and examine the average number of new instances of collaboration for authors who had a certain number of previous collaborations.

Figure 14 shows the logarithmic representation of the results. As demonstrated with the line of slope 1, the trend is very close to linear. However, there are two points, corresponding to $k = 7$ and 9, which are significantly below the others. What is happening there?

The probability of new instances of collaboration is obtained by dividing the number of new papers of authors who have collaborated $k$ times by the number of author pairs who have collaborated $k$ times. The author pair distribution of the previous instances of collaboration is generally a very regular power law (number of author pairs with an increasing number of previous collaborations drops like a power law), however, the number of author pairs who have collaborated 7 and 9 times is anomalously large. This anomaly is the result of 9 papers that contained in excess of 200 authors, most of which had the same authors on all 9 papers. Removing from the analysis all papers containing more than 200 authors, as in the previous analyses of preferential attachment, produces a much cleaner correlation (Figure 15), with a slope having an exponent of 1.03. This is the example of the analysis where hyper-authored papers can affect behaviors even at modest values of some variable, since the difference between Figure 14 and 15 lies solely at $k < 10$.

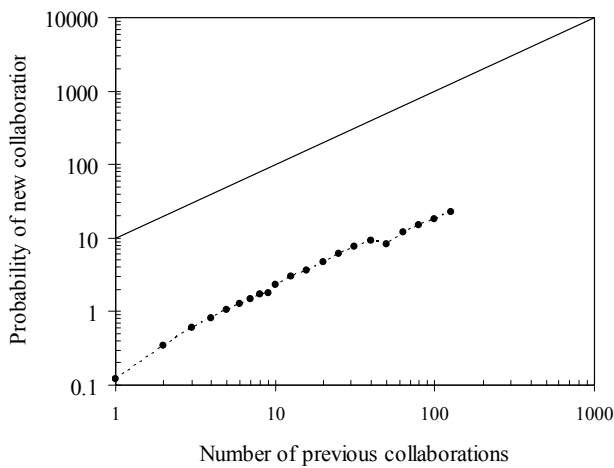

**Figure 15.** Probability of a new collaboration among the authors who have collaborated as a function of the number of previous collaborations with papers with 200 authors excluded. Linear preferential attachment is shown as a solid line.

*Collaboration within six representative fields*

Having established collaboration patterns for nanoscience as a whole, given its heterogeneity and assuming that collaboration practices differ in subfields that have different cultures and practices of publication and authorship, it would be instructive to examine collaboration within a select number of nanoscience subfields. To this end coauthorship networks divided by research subfields have been created. Subfield networks are subsets of full collaboration network, in that each subfield network contains all authors belonging to that subfield and their collaborators (which can belong to any subfield). Out of 42 subfields we will focus on six. These are: (1) Condensed Matter & Applied Physics (P2), (2) Analytical Chemistry (C2) and (3) Multidisciplinary Materials Science (M2) as the subfields most frequently assigned to articles (together accounting for 29% of all articles in 2000-04) plus (4) Nanoscience & Nanotechnology[7] (N1), (5) Social Sciences (SS1) as an example of a non-physical science, and (6) Physics of Particles and Fluids (P6), since particle physics (also called high energy physics) most often produces papers with hyperauthorship practices. In particular we focus on the distributions of the number of collaborators for these six subfields. The distributions (again for 2000-04) are shown in Figure 16 (a-f). Each panel contains the distribution of the entire network for comparison.

Condensed Matter & Applied Physics (P2), Analytical Chemistry (C2), Multidisciplinary Materials Science (M2) and Nanoscience & Nanotechnology (N1) exhibit log-normal behavior but with peaks occurring at different $k$ values. The peak is at four collaborators for P2 and N1 and at three for C2 and M2. Physics of Particles and Fluids (P6) shows a slight "hook" with the peak at two collaborators per author. In Social Sciences (SS1) the curve does not turn over for small $k$, i.e., most authors have one collaborator. All six subfields show power law parts with slopes that are similar to full networks, except P2 which is more shallow. C2 shows signs of an anomalous peak, while such a peak is dominating in P6, accounting for much of the total distribution in that range.

**Discussion and conclusions**

In a large-scale network study the distribution of the number of collaborators in the five-year period (2000-04) in nanoscience, an emerging highly interdisciplinary field, was examined to determine underlying processes with special attention to non-power law behavior and its causes.

The results indicate that there are three types of collaboration modes in nanoscience that correspond to three ranges in the distribution of collaborators, i.e., network node distribution (see Figure 1): (1) the log-normal "hook", (2) the power law tail, and (3) the anomalous peak in the power law. While most attention in the literature has been paid to the power law distribution and its origin, a large majority of authors actually belongs to non-power law "hook" regime (88% of collaborating nanotechnology authors).

---

[7] While all the articles studied here are considered to be nanotechnology-related, and all the authors are considered to be producing nanotechnology-related research, nanotechnology as a subfield consists of authors who predominantly publish their research in the journals whose main subject is nanotechnology (as defined by ISI subject categories of journals). Interestingly, only 3% of nanotechnology articles are published in exclusive nanotechnology journals (Milojević, 2009).

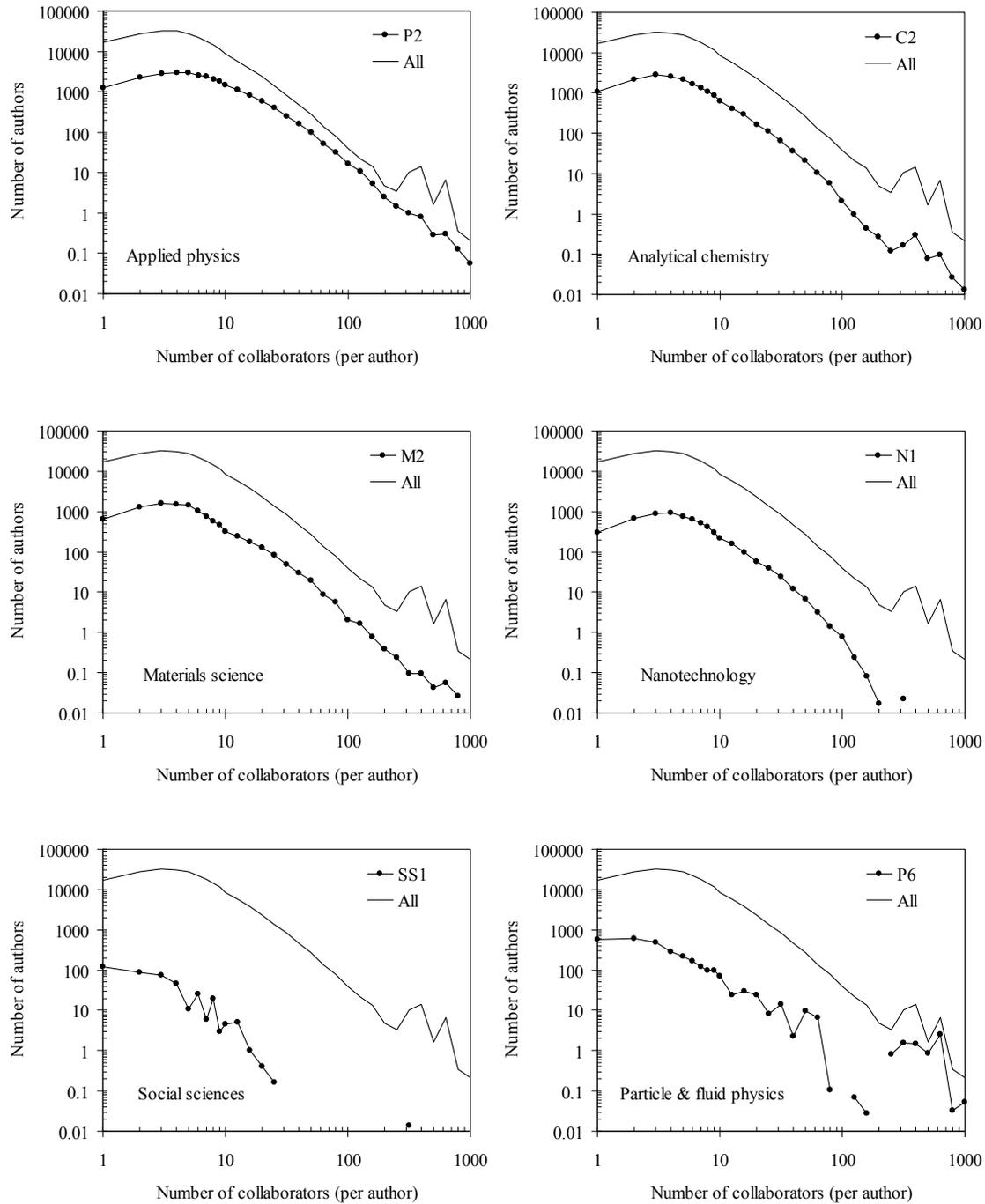

**Figure 16 (a-f). The distributions of the number of collaborators for 6 subfields (dotted line, P2, C2, M2, N1, SS1, P6) and the full distribution (line) for the period 2000-04. What is shown is the number of collaborators (from *any* subfield) of an author belonging to a *given* subfield.**

In their study of the growth of international collaboration in science, Wagner and Leydesdorff (2005) observed the existence of what they call "hooks" in the degree distributions of five out of six fields of science (they find hooks in Astrophysics, Polymer science, Geophysics, Soil science and Virology, but not in Mathematical Logic). They also point out that "similar deviations" were found in the degree distributions of fields studied by Newman (2001b). It must be noted that the type of deviation that Newman found is different from the "hook". Newman finds that part of the distribution involving authors with small number of collaborators does indeed follow a power law, but that the distribution for part of the distribution for authors having *more* than between 10 and 100 collaborators falls below the power law constrained by low $k$. To model this part of the distribution (large $k$) Newman introduces an exponential cut-off to the power law. Thus, hooks do not seem to be clearly present in distributions shown by Newman (his Figure 1), even for same fields later studied by Wagner and Leydesdorff. One possible explanation for this is that Newman's logarithmic binning scheme, which is applied to all $k$, somehow boosts points with small $k$ and washes out the hook. So the node degree distributions presented by Newman do not look qualitatively similar to those presented in this paper or to those shown by Barabási et al. (2002) (their Figure 2) or Wagner and Leydesdorff, which all generally show a hook for authors with small number of collaborators and the power law afterwards.

Despite the obvious presence of the hook in their data, few have tried to explain this phenomenon, instead focusing on the power law tail that actually contains a small number of authors in a collaboration network. An exception is the work of Wagner and Leydesdorff, who suggest a sociological explanation in which the "hook" is the consequence of the arrival of newcomers (scientists at the beginning of their careers) and transients (scientists who publish in the field only once) in the field. According to them, having scientists with different levels of seniority leads to two types of preferential attachment being at play, which then cause different behavior in the collaborator distribution (but note that they do not actually test for preferential attachment nor perform separate analyses for newcomers and transients).

In this work the presence of various types of preferential is tested directly. For "hook" part of the distribution we demonstrate that the preferential attachment does not operate (Figures 9 and 11). Rather, the results show that the probability of acquiring a new collaborator (whether it is an author joining the network or someone with whom there was no collaboration in the past) for authors with small number of collaborators is independent of the number of previous collaborators. Without the preferential attachment one no longer expects the collaborator distribution to follow a power law, and we see that it does not. Instead, it falls below the power-law trend established by the tail (but not yet necessarily forms a hook, i.e., a feature with a maximum at $k > 1$). The exact functional form that results from the absence of the preferential attachment can be obtained from theoretical network modeling and is beyond the scope of this work.

The reason why this part of the collaboration distribution does not feature preferential attachment may be, as suggested by Wagner and Leydesdorff (2005), because it is mainly populated with newcomers and transients (Figure 7). These scientists who are just entering the field are seeking (or are being recruited by) collaborators who are already established (and therefore usually already have many collaborators) rather than the ones who are newcomers like themselves, which is why they do not benefit from preferential attachment.

However, while confirming that the "hook" is mainly populated by newcomers and transients, we show that the collaborator distribution of senior authors also has a clear hook, i.e., this feature is not merely the result of the amount of time spent in the field (Figure 6). This leads us to the very cause of the hook, i.e., a feature that not only falls below the power law (which is the consequence of the absence of preferential attachment) but also has a peak at $k > 1$. For the "hook" to disappear it would be necessary that most authors have only one collaborator. However, apparently such authors are not as frequent as those who have several collaborators, even among the newcomers. This tells us that there is a "natural" or "characteristic" mode of collaboration which is essentially a variation of a normal distribution, and therefore has a *peak*. This mode reflects the fact that in most modern fields of science the majority of scientists tend to coauthor papers with a small number of collaborators, rather than work alone or in very large groups. Coauthoring a paper with fewer people is easier to manage. Since collaboration requires "investing" time and energy to "learn" to work with other researchers, it is not surprising that most scientists naturally tend to invest less time and energy, especially since even lower investment generates returns in terms of increased productivity.

These findings are further strengthened by the fact that the extent and peak of the "hook" is different from subfield to subfield (Figure 16 (a-f)), and as noted above may represent the typical "optimal" number of collaborators authors are willing to work with to generate a publication. Apparently, that number is larger in some nanotechnology subfields than in others and this is presumably a reflection of type of the research conducted in these original fields. In other words, various authors publishing nanotechnology articles must be retaining (to a degree that cannot be determined from these data alone) authorship and/or collaboration practices of their "native" fields, emphasizing the multidisciplinary nature of most nanotechnology work. Thus, the peak of the "hook" is larger in more experimental fields requiring the involvement of larger teams, and smaller in more theoretical ones. Of the subfields explored here it is only Social Sciences that lacks the "hook", and where consequently most authors have only one collaborator (but note that even there the collaboration distribution for small number of collaborators falls below the power law trend of those with larger $k$).. Wagner and Leydesdorff (2005) found a similar distribution for the field of Mathematical Logic, but they doubted the significance of that "exception" because of the small number of papers that were included in that field. On the contrary, we regard such distributions as genuine, and simply indicating a very low "characteristic" number of collaborators. Indeed, of the 42 subfields of nanotechnology we find only one other, Mathematics, where also most authors have a single collaborator.

Another indication that the "hook" part of the distribution belongs to a specific mode of collaboration is that the authors who only publish alone and therefore have zero collaborators form a natural and smooth extension of the log-normal distribution (Figure 8) with the number of isolates even smaller than those who have only one collaborator. Of the two nanotechnology subfields where collaboration distribution peaks for authors with a single collaborator, the number of isolates in Mathematics is smaller than the number of those with one collaborator, while in Social Sciences the number of isolates is actually two times higher than the number of authors with one collaborator. So the characteristic number of collaborators for Social Sciences authors publishing nanotechnology articles is zero, presumably a practice of the social sciences in general.

The results also indicate (Figure 6) that the range of $k$ where this normal mode of collaboration holds is wider for authors who have spent increasing number of years in the field. In other words, these authors obtain collaborators less through the process of preferential attachment (i.e., they are not "star" scientists) and more as the result of the extensive normal mode of collaboration.

The existence of power law is a reflection of the well-known existence of scientific stars, or high-status scientists who attract a disproportionate number of collaborators through the process of preferential attachment. Figure 9 confirms that new scientists are more likely to collaborate with scientists who already have large number of collaborators. Testing of the two scenarios for preferential attachment among the existing authors has shown that the probability of new collaborations does not depend on the number of mutual collaborators, but only on the total number of collaborators (i.e., follows the form of preferential attachment suggested by Barabási et al. (2002)). This might indicate that the status of scientists (as determined by the number of collaborators they have) is more important for initiating new collaborations than the existence of common collaborators. We have also established the formal presence of the preferential attachment among the authors who have collaborated previously, so that the probability of a repeat collaboration increases with the number of previous instances of collaboration. However, is this type of preferential attachment justifiably called "attachment"? This may depend on what is meant by "attachment". Unlike the previous two types, here there is no creation of new edges in the network, but rather the increase in weight of the existing edges (which are unimportant in most measures of network topology). Actually, the claim here can be restated that the weights of the edges in the new network will on average be proportional to the weights of the edges in old network. However, this may not be all that surprising because if one simply assumes that the authors who collaborated with some frequency in the past will on average continue to collaborate with the same frequency in the future, this will automatically lead to the probability of new instance of collaboration being proportional to the number of previous instances.

We have shown (Figure 4) that the authors who are in the "anomalous peak" follow hyperauthorship practices. Namely, the large number of collaborators for these authors is the result of them being part of large teams (200 authors or more) who publish papers together. Being team members these authors do not benefit individually from the processes of preferential attachment (Figures 9, 11 and 14). These results are another indicator (see also Newman 2004, p. 11) that one should be careful when using coauthorship networks as representations of collaboration in fields where a high number of coauthors per paper, or team authorship, is the norm. Considering such coauthors collaborators may be problematic, given that in large groups coauthors may not even know each other, or exchange information and interact professionally. However, the results from this study indicate that although there are areas of nanoscience exhibiting hyperauthorship practices, nanoscience as a whole is not such a field.

The majority of researchers has a fairly small number of collaborators, and in nanoscience this distribution has a log-normal form with a peak at three collaborators. These scientists follow a regime of collaboration that requires a certain optimal number of collaborators to carry out research, the exact number being dependent on the type of the field. These authors do not benefit from preferential attachment which is why the log-normal hook falls below the power law trend of authors with many collaborators. While there are many newcomers and transients in this category it also contains a number of continuing authors, who despite the length of time spent in the field do not have a "star" status. Beyond the log-normal part, there are scientists who obviously benefit from the processes of preferential attachment. These "star" scientists hold central positions having disproportionately large number of collaborators. And finally, authors who work in very large teams, following hyperauthorship practices, clearly do not benefit from preferential attachment. Using coauthorship in this way to study collaboration in these fields is not warranted.

The main limitation of this study is that while it identifies the processes that lead to the observed collaborator distribution, these processes have not been theoretically modeled in a dynamic way, i.e., no attempt has been made to evolve a network using a set of rules and reproduce the functional form of the collaborator distribution. This will be a goal of the forthcoming study.

**Acknowledgments**

I am grateful to Jonathan Furner, Blaise Cronin and Katy Börner for their comments and feedback. I would also like to thank Gregory Leazer, and Phillip Bonacich for useful discussions. Certain data included herein are derived from NanoBank (Lynne G. Zucker and Michael R. Darby, NanoBank Data Description, release 1.0 (beta-test), Los Angeles, CA: UCLA Center for International Science, Technology, and Cultural Policy and NanoBank, January 17, 2007.) Certain data included herein are derived from the Science Citation Index Expanded, Social Sciences Citation Index, and Arts & Humanities Citation Index of the Institute for Scientific Information®, Inc. (ISI®), Philadelphia, Pennsylvania, USA: © Copyright Institute for Scientific Information®, Inc. 2006. All rights reserved.

**References**

Allison, P. D., Krauze, T. K., & Long, J. S. (1982). Cumulative advantage and inequality in science. *American Sociological Review, 47*(5), 615-625.

Barabási, A.-L., & Albert, R. (1999). Emergence of scaling in random networks. *Science, 286*, 509-512.

Barabási, A.-L., Jeong, H., Neda, Z., Ravasz, E., Schubert, A., & Vicsek, T. (2002). Evolution of the social network of scientific collaborations. *Physica A, 311*, 590-614.

Barben, D., Fisher, E., Selin, C., & Guston, D. H. (2008). Anticipatory governance of nanotechnology: Foresight, engagement, and integration. In E. J. Hackett, O. Amsterdamska, M. Lynch & J. Wajcman (Eds.), *The Handbook of Science and Technology Studies* (3rd ed. ed., pp. 979-1000). Cambridge: MIT Press.

Beaver, D. d. (2001). Reflections on scientific collaboration (and its study): past, present, and future. *Scientometrics, 52*(3), 365-377.

Beaver, D. d., & Rosen, R. (1978). Studies in scientific collaboration. Part I. The professional origins of scientific co-authorship. *Scientometrics, 1*(1), 65-84.

Beaver, D. d., & Rosen, R. (1979a). Studies in scientific collaboration. Part II. Scientific co-authorship, research productivity and visibility in the French scientific elite, 179901830. *Scientometrics, 1*(2), 133-149.

Beaver, D. d., & Rosen, R. (1979b). Studies in scientific collaboration. Part III. Professionalization and the natural history of modern scientific co-authorship. *Scientometrics, 1*(3), 231-245.

Berube, D. M. (2006). *Nano-hype: The truth behind the nanotechnology buzz*. Amherst: Prometheus Books.

Birnholtz, J. P. (2006). What does it mean to be an author? The intersection of credit, contribution, and collaboration in science. *Journal of the American Society for Information Science and Technology, 57*(13), 1758-1770.

Bordons, M., & Gomez, I. (2000). Collaboration networks in science. In B. Cronin & H. B. Atkins (Eds.), *The web of knowledge: A festschrift in honor of Eugene Garfield* (pp. 197-213). Medford, NJ: Information Today.

Börner, K., Sanyal, S., & Vespignani, A. (2007). Network science. In B. Cronin (Ed.), *Annual Review of Information Science and Technology* (Vol. 41, pp. 537-607). Medford, NJ: Information Today.

Cole, J. R., & Cole, S. (1973). *Social stratification in science*. Chicago: University of Chicago Press.

Crane, D. (1972). *Invisible colleges: diffusion of knowledge in scientific communities*. Chicago: University of Chicago Press.

Cronin, B. (2001). Hyperauthorship: A postmodern perversion of evidence of a structural shift in scholarly communication practices? *Journal of the American Society for Information Science and Technology, 52*(7), 558-569.

Dorogovtsev, S. N., & Mendes, J. F. F. (2000). Scaling behaviour of developing and decaying networks. *Europhysics Letters, 52*, 33-39.

Dorogovtsev, S. N., & Mendes, J. F. F. (2002). Evolution of networks. *Advances in Physics, 51*(4), 1079-1187.

Farkas, I., Derenyi, I., Jeong, H., Neda, Z., Oltvai, Z. N., Ravasz, E., et al. (2002). Networks in life: Scaling properties and eigenvalue spectra. *Physica A, 314*(1-4), 25-34.

Glänzel, W. (2002). Coauthorship patterns and trends in the sciences (1980-1998): A bibliometric study with implications for database indexing and search strategies. *Library Trends, 50*(3), 461-473.

Glänzel, W., & Schubert, A. (2004). Analysing scientific networks through co-authorship. In H. F. Moed, W. Glänzel & U. Schmoch (Eds.), *Handbook of quantitative science and technology research: The use of publication and patent statistics in studies of S&T systems* (pp. 257-276). Dordrecht: Kluwer Academic Publishers.

Grodal, S. (2007). *The emergence of a new organizational field - lables, meaning and emotions in nanotechnology*. Stanford University, Palo Alto.

Hara, N., Solomon, P., Kim, S.-L., & Sonnenwald, D. H. (2003). An emerging view of scientific collaboration: Scientists' perspectives on collaboration and factors that impact collaboration. *Journal of the American Society for Information Science and Technology, 54*(10), 952-965.

Hayles, N. K. (2004). Connecting the quantum dots: Nanotechscience and culture. In N. K. Hayles (Ed.), *Nanoculture: Implications of the new technoscience* (pp. 11-26). Bristol: Intellect Books.

Katz, J. S., & Martin, B. R. (1997). What is research collaboration? *Research Policy, 26*, 1-18.

Kretschmer, H. (1997). Patterns of behaviour in coauthorship networks of invisible colleges. *Scientometrics, 40*(3), 579-591.

Lee, S., & Bozeman, B. (2005). The impact of research collaboration on scientific productivity. *Social Studies of Science, 35*(5), 673-702.

Luukkonen, T., Persson, O., & Sivertsen, G. (1992). Understanding patterns of international scientific collaboration. *Science, Technology & Human Values, 17*(1), 101-126.

Melin, G., & Persson, O. (1996). Studying research collaboration using co-authorship. *Scientometrics, 36*(3), 363-377.

Merton, R. K. (1968). The Matthew effect in science. *Science, 159*(3810), 56-63.

Merton, R. K. (1988). The Matthew effect in science, II: Cumulative advantage and the symbolism of intellectual property. *ISIS, 79*, 606-623.

Milburn, C. (2004). Nanotechnology in the age of posthuman engineering: Science fiction as science. In N. K. Hayles (Ed.), *Nanoculture: Implications of the new technoscience* (pp. 109-129). Portland: Intellect Books.

Milojević, S. (2009). *Big Science, Nano Science? : Mapping the Evolution and Socio-Cognitive Structure of Nanoscience/nanotechnology Using Mixed Methods*. University of California Los Angeles, Los Angeles.

Moody, J. (2004). The structure of a social science collaboration network: Disciplinary cohesion from 1963 to 1999. *American Sociological Review, 69*(2), 213-238.

Newman, M. E. J. (2001a). Clustering and preferential attachment in growing networks. *Physical Review E, 64*(2), 025102(025104).

Newman, M. E. J. (2001b). The structure of scientific collaboration networks. *PNAS, 98*(2), 404-409.

Newman, M. E. J. (2003). The structure and function of complex networks. *SIAM Review, 45*(2), 167-- 256.

Newman, M. E. J. (2004). Who is the best connected scientist? A study of scientific coauthorship networks. In E. Ben-Naim, H. Frauenfelder & Z. Toroczkai (Eds.), *Complex networks* (pp. 337–370). Berlin: Springer.

Price, D. J. d. S. (1976). A general theory of bibliometric and other cumulative advantage processes. *Journal of the American Society for Information Science, 27*(5), 292-306.

Shrum, W., Genuth, J., & Chompalov, I. (2007). *Structures of Scientific Collaboration*. Cambridge: MIT Press.

Smith, D., & Katz, J. S. (2000). *Collaborative Approaches to Research: Final Report*.

Sonnenwald, D. H. (2007). Scientific collaboration: Challenges and solutions. In B. Cronin (Ed.), *Annual Review of Information Science & Technology* (Vol. 41, pp. 643-681). Medford, NJ: Information Today.

Subramanyam, K. (1983). Bibliometric studies of research collaboration: A review. *Journal of Information Science, 6*, 33-38.

Wagner, C. S. (2008). *The New Invisible College: Science for Development*. Washington, DC: Brookings Institution Press.

Wagner, C. S., & Leydesdorff, L. (2005). Network structure, self-organization, and the growth of international collaboration in science. *Research Policy, 34*(10), 1608-1618.

Wajcman, J. (2008). Emergent technosciences. In E. J. Hackett, O. Amsterdamska, M. Lynch & J. Wajcman (Eds.), *The Handbook of Science and Technology Studies* (3rd ed. ed., pp. 813-816). Cambridge: MIT Press.

Wasserman, S., & Faust, K. (1994). *Social network analysis: Methods and applications*. Cambridge: Cambridge University Press.

Watts, D. J. (2004a). The "new" science of networks. *Annual Review of Sociology, 30*, 243-270.

Zucker, L. G., & Darby, M. R. (2005). Socio-economic impact of nanoscale science: Initial results and NanoBank. *NBER Working Paper 11181*.

Zucker, L.G., & Darby, M.R. (2007). Nanobank Data Description release 2.0 (beta-test). Los Angeles, CA: UCLA Center for International Science, Technology and Cultural Policy and Nanobank. January 17, 2007-February 2, 2009.

Zucker, L. G., Darby, M. R., Furner, J., Liu, R. C., & Ma, H. (2007). Minerva unbound: Knowledge stocks, knowledge flows and new knowledge production. *Research Policy, 36*(6), 850-863.

Zuckerman, H. (1977). *Scientific elite: Nobel Laureates in the United States*. New York: The Free Press.